\documentclass[11pt]{article}
\usepackage{amsmath,amssymb,amsfonts}
\usepackage{cite,graphicx,color}
\usepackage[colorlinks=true, linkcolor=blue, citecolor=blue, linktoc=all]{hyperref}
\usepackage[usenames,dvipsnames]{xcolor}
\usepackage[mathscr]{euscript}

\usepackage{nicefrac}

\DeclareMathAlphabet{\mathpzc}{OT1}{pzc}{m}{it}

\begin{document}
\newcommand{\be}{\begin{equation}}
\newcommand{\ee}{\end{equation}}
\newcommand{\beq}{\begin{eqnarray}}
\newcommand{\eeq}{\end{eqnarray}}
\newcommand{\bea}{\begin{eqnarray}}
\newcommand{\eea}{\end{eqnarray}}
\newcommand{\beqn}{\begin{eqnarray}}
\newcommand{\eeqn}{\end{eqnarray}}

\def\pa{\partial}
\newcommand{\un}[1]{\underline{#1}}
\newcommand{\ack}[1]{[{\bf Pfft!: {#1}}]}
\newcommand{\ackm}[1]{\marginnote{\small Pfft!: #1}}
\newcommand{\hlt}[1]{{\color{WildStrawberry}{\em #1}}\index{#1}}

\newcommand{\Dt}{{e}}
\newcommand{\acc}{A}
\newcommand{\LCGamma}{\mathring{\Gamma}}
\newcommand{\LCgamma}{\mathring{\gamma}}
\newcommand{\LCnabla}{\mathring{\nabla}}
\newcommand{\LCtnabla}{\mathring{\tilde{\nabla}}}

\newcommand{\thistitle}{
	Carroll Structures, Null Geometry and 
	Conformal Isometries
	}
\newcommand{\addresspi}{
	Perimeter Institute for Theoretical Physics, 
	31 Caroline St. N., Waterloo ON, N2L 2Y5, Canada
	}
\newcommand{\addressuiuc}{
	Department of Physics, University of Illinois,
 	1110 West Green St., Urbana IL 61801, U.S.A.
	}
\newcommand{\addresspoly}{
	CPHT, CNRS,
	Ecole Polytechnique, IP Paris,\footnote{Centre de Physique Th\' eorique, Centre National de la Recherche Scientifique, Institut Polytechnique de Paris.}
	91128 Palaiseau Cedex, France
	}
\newcommand{\emailrgl}{rgleigh@illinois.edu}
\newcommand{\emaillc}{ciambelli.luca@gmail.com}

\title{\thistitle}
\author{Luca Ciambelli,$^{a,b}$ Robert G. Leigh,$^{b,c}$ Charles Marteau$^a$ and P. Marios Petropoulos$^a$\\
	\\
	{\small ${}^a$\emph{\addresspoly}}\\ 
	{\small ${}^b$\emph{\addresspi}}\\
	{\small ${}^c$\emph{\addressuiuc}}\\ 
	\\
	}
\date{\today}
\maketitle
\vspace{-5ex}

\begin{abstract}
We study the concept of Carrollian spacetime starting from its underlying fiber-bundle structure. The latter admits an Ehresmann connection, which enables a natural separation of time and space, preserved by the subset of Carrollian diffeomorphisms. These allow for the definition of Carrollian tensors and the structure at hand provides the designated tools for describing the geometry of  null hypersurfaces embedded in Lorentzian manifolds. Using these tools, we investigate the conformal isometries of general Carrollian spacetimes and their relationship with the BMS group.
\end{abstract}

\setcounter{footnote}{0}
\renewcommand{\thefootnote}{\arabic{footnote}}

\section{Introduction}

The Carroll group was discovered by L\'evy-Leblond in 1965 \cite{Levy1965} as a dual  contraction of the Poincar\'e group, operating at vanishing rather than infinite velocity of light.  The increasing interest in non-Minkowskian spacetimes possessing nonetheless boost-like isometries, has led to more systematic studies of Carrollian constructions. Besides the intrinsic value of the latter (along with Newton--Cartan), the resurgence in the area has been sustained by the parallel growth of two distinct albeit related fields of application. The first involves codimension-one null hypersurfaces in Lorentzian  \emph{i.e.},  hyperbolic pseudo-Riemannian manifolds. 
The second concerns the development of flat holography.

Carroll structures were introduced in \cite{Duval:2014uoa, Duval:2014uva, Duval:2014lpa} as alternatives to Riemannian or Newton--Cartan geometries.\footnote{See also the earlier publication \cite{Henneaux:1979vn} on geometries with degenerate metric.} According to these authors, Carroll structures consist of a $d+1$-dimensional manifold $\mathscr{C}$ equipped with a degenerate metric $g$ and a vector field $E$, which defines the kernel of the metric, \emph{i.e.} $g(E,\ldotp)=0$. In this definition, the Carroll group
emerges as the isometry group of \emph{flat} Carrollian structures, whereas general diffeomorphisms are always available. Because of the field $E$, the Carroll structure defines a natural separation between time and space, and a subset of diffeomorphisms arises, the Carrollian diffeomorphisms, which preserves this separation.  

Given their defining properties, Carroll structures are expected to arise systematically  as  geometries on null hypersurfaces of relativistic spacetimes, because the induced metric inherited from the embedding is degenerate (see \emph{e.g.} \cite{Hartong:2015xda}).
There are several notable instances of null hypersurfaces. Generally, null hypersurfaces occur as components of the  boundary of causal diamonds and related structures, relevant in the study of entanglement. One also finds null hypersurfaces in other important physical situations, such as black-hole horizons and the hypersurfaces appearing at light-like infinity of asymptotically flat spacetimes (commonly designated as $\mathscr{I}^{\pm}$).  The latter makes the bridge with asymptotically flat holography, in which the putative dual degrees of freedom are expected to be defined precisely on this null-infinity hypersurface. In fact, asymptotically flat holography has been probably the first arena of application of Carrollian physics \cite{Barnich:2010eb, Bagchi2010e}, not so much because of the geometric structure the boundary is endowed with (its Carrollian nature was identified much later), but for the emergence of the BMS symmetry. The BMS group was discovered in 1962 by Bondi, van der Burg, Metzner and Sachs \cite{Bondi1962, Sachs1962} as the asymptotic isometry group of asymptotically flat  spacetimes towards light infinity, and became popular lately in relation with null hypersurfaces and flat holography (see \emph{e.g.} \cite{Barnich:2009se, Ashtekar:2014zsa}).  
It was in particular proven \cite{Duval:2014uva} that the $\mathfrak{bms}(d+2)$ algebra is isomorphic to the conformal Carroll algebra $\mathfrak{ccarr}(d+1)$ for $d=1,2$. This is yet another sign corroborating the triangle ``Carroll--null--BMS''.

The aim of the present work is to revisit this web of relationships and provide an alternative perspective to some of its aspects. Our analysis follows two paths. On the one hand, we define a Carrollian spacetime in terms of a fiber bundle accompanied with a Carroll structure. The ingredients are thus an Ehresmann connection, a degenerate metric and a scale factor,\footnote{Note that these ingredients all appear within the context of Carrollian fluids and the fluid-gravity correspondence, as in Refs. \cite{Ciambelli:2018xat, Ciambelli:2018wre}.} all assumed \emph{a priori} time- and space-dependent. This
provides us with a geometric understanding of the appearance of Carrollian diffeomorphisms and the reduction of spacetime tensors to Carrollian 
tensors. Carrollian spacetimes with the above set of ingredients are also naturally revealed in null embedded hypersurfaces. 
On the other hand, we discuss the conformal isometry algebra of general Carrollian spacetimes.  In the \emph{shearless} case (properly defined shortly), we generally recover the familiar algebra of transformations. In two and three dimensions, the algebra coincides with BMS, whereas in arbitrary dimension it appears as the semi-direct product of the conformal isometry group of the metric with supertranslations. The strength of our results resides in their wide validity for shearless but otherwise arbitrary Carrollian geometries. In the literature there have been other proposals made for a notion of geometry defined on null embedded hypersurfaces, the ``universal structures'', (see \emph{e.g.}, \cite{Chandrasekaran:2018aop}). Different such proposals may lead to different algebras that preserve the given structure, with subsequently a potential choice of partial gauge fixing.

\section{Carrollian Spacetimes as Fiber Bundles} \label{carfib}

\subsubsection*{The Intrinsic Definition}

A $d+1$-dimensional Carrollian spacetime $\mathscr{C}$ is elegantly described in terms of a fiber bundle, with one-dimensional fibers, and a $d$-dimensional base $\mathscr{S}$ thought of as the space, the fiber being the time. As usual, the bundle structure provides a projection  $\pi:\mathscr{C}\rightarrow\mathscr{S}$, which defines in turn a surjective linear map between the corresponding tangent bundles, $\text{d}\pi:T\mathscr{C}\rightarrow T\mathscr{S}$. It is convenient to choose a local coordinate system $x=\{t,\mathbf{x}\}$ such that the action of the projector simplifies to $\pi:(t,\mathbf{x})\rightarrow\mathbf{x}$, that is, $t$ is the fiber coordinate. 

One can define a vertical subbundle as $V=\text{ker}(\text{d}\pi)$. The above coordinate set has been chosen such that $V$ is given by all sections of $T\mathscr{C}$ proportional to $\partial_t$ (vectors of the vertical tangent subspace $V_{(t,\mathbf{x})}$ are of the form $W^t\partial_t$).
In order to split the tangent space $T_{(t,\mathbf{x})}\mathscr{C}$ into a direct sum of vertical and horizontal components, $V_{(t,\mathbf{x})} \oplus H_{(t,\mathbf{x})}$, smooth everywhere \emph{i.e.} valid for the tangent bundle, $T\mathscr{C}=V
\oplus H$, one needs an \hlt{Ehresmann connection.} With this connection, the linear map $\text{d}\pi$ restricted to $H_{(t,\mathbf{x})}$ sets a one-to-one correspondence between $H_{(t,\mathbf{x})}$ and  $T_{\mathbf{x}}\mathscr{S}$. This allows to lift vertically vectors 
 $W=W^i\pa_i\in T_{\mathbf{x}}\mathscr{S}$ to $\bar W=W^iE_i\in H_{(t,\mathbf{x})}$,
where 
\begin{equation}
\label{Di}
E_i=\partial_i+b_i\partial_t, \quad i=1,\ldots,d
\end{equation} 
provide a  basis for $H_{(t,\mathbf{x})}$. The Ehresmann connection is encoded in the 
one-form field $\pmb{b}=b_i(t,\mathbf{x})\text{d}x^i \in T^*\mathscr{C}$.

The Ehresmann connection has many facets. On the one hand, it provides a lift of curves in $\mathscr{S}$ onto curves in $\mathscr{C}$ such that the tangent vectors to the latter are horizontal.
On the other hand, it makes it possible to realize the splitting $T\mathscr{C}=V\oplus H$ through the definition of the projector $p$ acting on $T\mathscr{C}$ with image $V$ and kernel $H$:
\begin{equation}
p=\partial_t\otimes (\text{d}t-b_i(t,\mathbf{x})\text{d}x^i).
\label{Ehres}
\end{equation} 

We will call the fiber bundle $\mathscr{C}$ a Carrollian spacetime, once endowed with a degenerate metric $g$ whose one-dimensional kernel coincides with the vertical subbundle $V$:
\begin{equation}
g(X,\ldotp)=0, \quad\forall X\in V.
\end{equation}
In the local coordinate system this imposes the metric be of the form
\begin{equation}
\label{degg}
g=g_{ij}(t,\mathbf{x})\,\text{d}x^i\otimes\text{d}x^j.
\end{equation}
providing a time-dependent notion of distances.

At this point of the presentation, it is worth mentioning that the triple $(\mathscr{C},V,g)$ corresponds to the definition of a \emph{weak} Carrollian structure given in \cite{Duval:2014uoa}. Together with this triple, the Ehresmann connection defines a Leibnizian structure \cite{Bernal2003, Bekaert:2014bwa, Bekaert:2015xua}. From the spacetime viewpoint, the fiber-bundle structure and the accompanying Ehresmann connection are the key ingredients for the intrinsic horizontal versus vertical splitting of the tangent bundle, and more generally of any tensor bundle. 

The coordinate system $\{t,\mathbf{x}\}$ is adapted to the splitting at hand, as is any new chart obtained through the transformation    
\begin{equation}
\label{cardiff}
t\mapsto t'(t,\mathbf{x})\quad\text{and}\quad \mathbf{x}\mapsto\mathbf{x}'(\mathbf{x}).
\end{equation}
The motivation for introducing the fiber-bundle structure is, among others, to make these diffeomorphisms natural, being a reparameterization of the fiber coordinate at each spatial point and a change of coordinates on the base, respectively.
With this, the Jacobian matrix $J^\mu_\nu=\frac{\pa x^{\prime\mu}}{\pa x^\nu}$  is upper triangular:
\beq
\begin{pmatrix}
J(t,\mathbf x) & J_i(t,\mathbf x)\\ 0 & J_i^j(\mathbf x)
\end{pmatrix},
\label{jac}
\eeq
since
\beq
\text{d}t'=J(t,\mathbf x)\text{d}t+J_i(t,\mathbf x)\text{d}x^i,\quad \text{d}x^{\prime j}=J_i^j(\mathbf x)\text{d}x^i,\label{FC}
\eeq
or equivalently  
\beq
\pa_{t}=J(t,\mathbf x)\pa_{t}^{\prime},\quad
\pa_{i}=J_i(t,\mathbf x)\pa_{t}^{\prime}+J^j_i(\mathbf x)\pa_{j}^{\prime }.
\eeq
These diffeomorphisms were called \emph{Carrollian} in \cite{Ciambelli:2018xat}. Every spacetime tensor field can be decomposed intrinsically into vertical and horizontal components, the latter transforming tensorially under Carrollian diffeomorphisms. These components are the \emph{Carrollian tensors} introduced in \cite{Ciambelli:2018xat}. An example of Carrollian tensor is  the degenerate metric \eqref{degg}, whose components transform as 
\begin{equation}
\label{gtr}
g'_{ij}=J^{-1k}_{\hphantom{-1}i}J^{-1\ell}_{\hphantom{-1}j}g_{k\ell}
\end{equation}
\emph{i.e.}, as a rank-$(0,2)$ Carrollian tensor field. In order to maintain $p$ in Eq.  \eqref{Ehres} invariant, the components of the Ehresmann connection must transform as:
\begin{equation}
\label{btr}
b'_j=J^{-1i}_{\hphantom{-1}j}\left(Jb_i+J_i\right).
\end{equation}

For reasons that will become clear in the course of the paper, it is convenient to introduce a density $\Omega(t,\mathbf x)$, transforming under Carrollian diffeomorphisms as:\footnote{Observe that 
$\pmb{b}$ transforms as a Carrollian connection density. Strictly speaking, the Ehresmann connection is thus
$\pmb{b}\Omega$. To avoid confusion, we should mention that the latter combination was  precisely used as $\pmb{b}$ in \cite{Ciambelli:2018xat, Ciambelli:2018wre}.
}
\begin{equation}
\label{Omtr}
\Omega'(t',\mathbf x')=J(t,\mathbf x)^{-1}\Omega(t,\mathbf x).
\end{equation}
With this density, one defines a new basis vector of $V_{(t,\mathbf{x})}$ as 
\begin{equation}
\label{Dt}
E=\Omega(t,\mathbf{x})^{-1}\partial_t.
\end{equation}
Together with the $H_{(t,\mathbf{x})}$ basis vectors $E_i$ defined in \eqref{Di}, we obtain a frame $E_\mu, \mu=0,\ldots, d$, adapted to the split tangent space and transforming canonically under Carrollian diffeomorphisms ($E_0\equiv E$):
\begin{equation}
E'=E\quad\text{and}\quad E_i'=J^{-1j}_{\hphantom{-1}i}E_j.
\end{equation}
The dual coframe, generically referred to as $\pmb{e}^\mu, \mu=0,\ldots, d$, is ($\pmb{e}^0\equiv \pmb{e}$)  
\begin{equation}
\label{cofD}
\pmb{e}=\Omega\left(\text{d}t-b_j\text{d}x^j\right) \quad \text{and}\quad \pmb{e}^i=\text{d}x^i,\quad  i=1,\ldots, d
\end{equation}
with $\pmb{e}^i$ transforming as in \eqref{FC} and $\pmb{e}'=\pmb{e}$. 

Any vector $W\in T\mathscr{C}$ is decomposed in the above frame as $W=W^0(t,\mathbf x) E +W^i(t,\mathbf  x) E_i$, while any form $\pmb{\omega}\in T^\ast  \mathscr{C}$
is $\pmb{\omega}=\omega_0(t,\mathbf x) \pmb{e} +\omega_i(t,\mathbf  x) \pmb{e}^i$. In this basis, the vertical and horizontal components are reduced, \emph{i.e.}, do not mix under Carrollian diffeomorphisms. The vertical components remain invariant, while the horizontal transform tensorially under Carroll diffeomorphisms:
\beq
W^{\prime 0}=W^0,\quad W^{\prime i}=J^i_j W^j,\quad \omega^{\prime}_{0}=\omega_0,\quad \omega^{\prime }_{i}=J^{-1j}_{\hphantom{-1}i} \omega_j.
\eeq
From the horizontal perspective $W^0$ and $\omega_0$ are scalars, and we refer to them as \emph{Carrollian scalars}, whereas $W^i$ and $\omega_i$ are components of a \emph{Carrollian vector} and a \emph{Carrollian one-form}. The same reduction properties are valid for rank-$(r,s)$ tensor fields in $T^{(r,s)}\mathscr{C}$. Notice that one can use $g_{ij}=g(E_i,E_j)$ and its inverse $g^{ij}$ for lowering and raising spatial indices $i,j,\ldots$ amongst Carrollian tensors.

In terms of the frame \eqref{Di}, \eqref{Dt}, and the coframe \eqref{cofD}, the action of the exterior derivative on the generic one form $\pmb \omega$ reads:
\begin{equation}
\text{d}\pmb \omega=(E(\omega_i)-E_i(\omega_0))\pmb{e}\wedge \pmb e^i+E_k(\omega_i)\pmb{e}^k\wedge \pmb e^i.
\end{equation}

One can define the Ehresmann curvature as 
\begin{equation}
\text{d}\pmb{e}=\pmb{\varphi}\wedge\pmb{e}+\pmb\varpi=
\varphi_i\, \pmb{e}^i\wedge \pmb{e}+\tfrac12\varpi_{ij}\, \pmb{e}^i\wedge \pmb{e}^j,
\label{db}
\end{equation}
which exhibits a pair of genuine Carrollian tensors.
The purely horizontal piece $\pmb\varpi$ is a Carrollian two-form, which we will call the \hlt{Carrollian torsion}.\footnote{The quantity $-\frac{1}{2}\varpi_{ij}$ is also referred to as  the Carrollian vorticity of the vector field $E$ \cite{Ciambelli:2018xat}.}
It has components 
\begin{equation}
 \varpi_{ij}=-\Omega\left(E_i(b_j)-E_j(b_i)\right).
 \label{torsion}
\end{equation}
The vertical-horizontal mixed components 
\beq
\label{varphi}
\varphi_i= \Omega E(b_i)+E_i(\ln \Omega),
\eeq
define a Carrollian one-form $\pmb{\varphi}$, the \hlt{acceleration.} Both appear in the Lie bracket of the basis vectors:
\beq
\left[E_i, E_j\right]=-\varpi_{ij} E,\quad
\left[E_i, E\right]=-\varphi_i E,
\eeq
which is dual to \eqref{db}.

A natural question to ask is whether $H$ can be thought of as the tangent bundle of codimension-one hypersurfaces in $\mathscr{C}$. If this holds, $\mathscr{C}$ is foliated by a family of hypersurfaces  modeled on $\mathscr{S}$. This is indeed possible whenever $H$ is an \hlt{integrable distribution} in $T\mathscr{C}$. The corresponding integrability condition originates from Fr\"obenius' theorem stating that the Lie bracket of horizontal vectors must be horizontal, or equivalently, that the vorticity of the normal (vertical) vector should vanish: $\varpi_{ij}=0$. In other words, the Ehresmann curvature should have no  horizontal component. 

Besides the above Carrollian tensors emanating from the Ehresmann connection, others can be defined using the metric $g$. Those are of two kinds.  
\begin{enumerate}

\item The first is based on first time-derivatives (the metric components are generically functions of both $t$ and $\mathbf{x}$): 
\beq
\label{thetashe}
\theta=E\left(\ln\sqrt{\det g}\right)=\frac12 g^{ik}E(g_{ki}),\quad
\zeta_{ij}=\frac12 E(g_{ij})-\frac{\theta}{d} g_{ij},
\eeq
referred to as \hlt{expansion} and \hlt{shear} ($g^{ij}$ are the components of the inverse of $g$). They are respectively a Carrollian scalar and a Carrollian symmetric and traceless rank-two tensor. The latter vanishes if and only if the time dependence in the metric is factorized:
$g_{ij}(t,\mathbf x)=\text{e}^{2\sigma(t,\mathbf x)}\tilde g_{ij}(\mathbf x)$,
in which case the expansion reads $\theta=d\, E(\sigma)$. This instance will turn out to play a significant role later in the discussion of BMS symmetry (Sec. \ref{BMS}). 

\item The second class is second-order in derivatives, and corresponds to the curvature of a generalized Levi--Civita connection. This is a canonical connection, which defines a horizontal parallel transport \emph{i.e.} a  covariant derivative acting on Carrollian tensors and producing new Carrollian tensors. It was introduced in \cite{Ciambelli:2018xat} as $\text{D}=\text{E}+{\gamma}
$, dubbed  Levi--Civita--Carroll, 
with ${\gamma}$ the Christoffel--Carroll symbols:
\begin{equation}
\label{dgammaCar}
\gamma^i_{jk}(t,\mathbf  x)=\frac{1}{2}g^{il}\left(
E_j
(g_{lk})+E_k  (g_{lj})-
E_l (g_{jk})\right)=\Gamma^i_{jk}+c^i_{jk},
\end{equation}
where $\Gamma^i_{jk}$ are the ordinary  Christoffel symbols and
\begin{equation}
\label{cCar}
c^i_{jk}(t,\mathbf  x)=\frac{\Omega}{2}g^{il}\left(
b_jE(
g_{lk})+b_k E( g_{lj})-
b_lE( g_{jk})\right).
\end{equation}
This connection, also cast as $\text{D}=\nabla + \text{c}$ with $\nabla$ the Levi--Civita connection, 
is metric-compatible ($D_k g_{ij}=0$), and since $\gamma^i_{[jk]}=0$, the torsion is exclusively encoded in the commutator of $E_i$'s, \emph{i.e.} in $\varpi_{ij}$. Its curvature tensors can be worked out following \cite{Ciambelli:2018xat}. As opposed to the ordinary Levi--Civita connection for Riemannian manifolds, the Levi--Civita--Carroll is not the unique metric-compatible and torsionless connection one can define on $T\mathscr{C}$. This question has been addressed \emph{e.g.} in \cite{Bekaert:2014bwa, Bekaert:2015xua}. 

\end{enumerate}

In the spirit of \cite{Duval:2014uoa, Duval:2014uva, Duval:2014lpa}, one can introduce the concept of \emph{flat Carrollian spacetime} given in an adapted coordinate system by
\beq\label{stflcar}
g_{ij}=\delta_{ij},\quad \Omega=1,\quad b_i=\text{const.}
\eeq
For this case, the Ehresmann curvature $\varpi_{ij}$ as well as the acceleration $\varphi_i$, the shear $\zeta_{ij}$ and the expansion $\theta$   vanish, as do the Christoffel--Carroll symbols written above. Carrollian flatness implies the Ehresmann connection being a pure gauge.

\subsubsection*{Realization on Null Hypersurfaces}

We would like now to discuss the appearance of the above structures on null hypersurfaces $\mathscr{C}$ of  a Lorentzian spacetime $\mathscr{M}$. The pull-back $g$ of the ambient metric on null hypersurfaces is degenerate with one-dimensional tangent subbundle kernel $V$, and from this perspective the Carrollian structure encompassed in the triple $(\mathscr{C},V,g)$ emerges naturally. This feature has been discussed by several authors, the more complete account being in the already quoted Ref. \cite{Hartong:2015xda}. Our fiber bundle with Ehresmann connection approach, which is designed for separating explicitly Carrollian time and space, emerges naturally in null embeddings.  This requires appropriate gauge-fixing in the ambient Lorentzian spacetime.\footnote{See \cite{Speranza2019} for a recent discussion on foliations and symmetries that preserve them.}

We will illustrate the above in the case of a $d+2$-dimensional spacetime $\mathscr{M}$ foliated with null hypersurfaces. In this case the ambient metric reads
\begin{equation}
\label{ambmet}
\text{d}s^2_{\mathscr{M}}=g_{ab}\text{d}x^a\text{d}x^b=-2\Omega\Xi\left(\text{d}t-b_i\text{d}x^i+\theta^t\text{d}r-b_i \theta^i \text{d}r\right)\text{d}r+g_{ij}\left(\text{d}x^i+\theta^i\text{d}r\right)\left(\text{d}x^j+\theta^j\text{d}r\right),
\end{equation}
where $\Omega$, $\Xi$, $b_i$, $\theta^t$, $\theta^i$ and $g_{ij}$ depend on all the coordinates $(r,t,\mathbf{x})$ and $t$ is a retarded time. The constant-$r$ leaves of the foliation $\mathscr{C}_r$ define $d+1$-dimensional null hypersurfaces because the pull-back of the metric, $g_r=g_{ij}(r,t,\mathbf{x})\text{d}x^i\text{d}x^j$, is indeed degenerate.
The diffeomorphisms that preserve the form of this metric are 
\begin{equation}
r\mapsto r'(r),\quad t\mapsto t'(r,t,\mathbf{x}),\quad \mathbf{x} \mapsto \mathbf{x}'(r,\mathbf{x}).
\label{BulkDiff}
\end{equation}
Defining as usual
\beq
J^a_b=\frac{\partial x'^a}{\partial x^b},
\eeq
the various quantities involved transform as
\beqn
\Omega' &=& \left(J_t^t\right)^{-1} \Omega \\
b'_{j} &=& J^{-1i}_{\hphantom{-1}j} \left(J^t_t b_{i}+J^t_i\right)\\
g'_{ij}&=&J^{-1k}_{\hphantom{-1}i}J^{-1\ell}_{\hphantom{-1}j}g_{k\ell}\\
\Xi' &=&  \left(J_r^r\right)^{-1} \Xi\\
\theta'^t &=&  \left(J_r^t\right)^{-1}\left(J^t_t \theta^t-J^t_r+J^t_i\theta^i \right)\\
\theta'^i &=&  \left(J^t_r\right)^{-1}\left(J^i_j\theta^j-J^i_r\right).
\eeqn
Therefore, we see that $\Omega$, $b_i$ and $g_{ij}$ transform on every leaf as they do on a Carrollian spacetime, eqs. \eqref{Omtr}, \eqref{btr}, and \eqref{gtr}. Hence, the diffeomorphisms \eqref{BulkDiff} are interpreted as Carroll diffeomorphisms on each leaf $\mathscr{C}_r$. The other elements $\Xi$, $\theta^t$ and $\theta^i$ were not present in the intrinsic definition of the previous section. This is not surprising as they account for the non-trivial $r$-dependence of the residual gauge symmetry \eqref{BulkDiff}. For simplicity we will fix locally $\Xi=1$ and $\theta^t=\theta^i=0$. This is achievable using (and therefore fixing) the $r$-dependence of the diffeomorphism \eqref{BulkDiff}. Henceforth the bulk metric simplifies to
\begin{equation}
\text{d}s^2_{\mathscr{M}}=-2\Omega\left(\text{d}t-b_i\text{d}x^i\right)\text{d}r+g_{ij}\text{d}x^i\text{d}x^j,
\end{equation}
with the residual gauge freedom \eqref{cardiff}:
\begin{equation}
r\mapsto r,\quad t\mapsto t'(t,\mathbf{x}), \quad \mathbf{x}\mapsto \mathbf{x}'(\mathbf{x}).
\label{BulkDiff2}
\end{equation}
Indeed, if we were to describe a single null hypersurface, it would also be natural to set, $\Xi=1$, and $\theta^i$ and $\theta^t$ to zero in its neighborhood. Under the coordinate change \eqref{BulkDiff2}, $g_{ij}$, $b_i$ and $\Omega$ still transform according to  \eqref{Omtr}, \eqref{btr}, and \eqref{gtr}. One can show that $\mathscr{C}_r$ equipped with these data is a $d+1$-dimensional Carrollian spacetime, in the lines we have discussed earlier. For this we need to exhibit the Ehresmann connection.

The ambient metric \eqref{ambmet} allows to define two independent null vector fields,\footnote{Our choice of gauge fixing differs from other works as \cite{Hartong:2015xda, Hopfmuller:2016scf}.} sections of $T\mathscr{M}$:
\begin{equation}
\ell=\frac{1}{\Omega}\partial_t, \quad n=\partial_r,\quad \ell\cdot n=1.
\label{}
\end{equation}
The corresponding forms in $T^\ast\mathscr{M}$ are
\begin{equation}
\pmb{\ell}=-\text{d}r, \quad \pmb{n}=\Omega\left(\text{d}t-b_i\text{d}x^i\right).
\label{}
\end{equation}
Hence, the vector field $\ell$ is normal to $\mathscr{C}_r$. Since it is null, it is also tangent to  $\mathscr{C}_r$ and belongs therefore to $T\mathscr{C}_r$. Being the kernel of the degenerate metric $g_r$ on $\mathscr{C}_r$, it spans the vertical subbundle $V_r$. The horizontal subbundle $H_r$ is given by the set of vectors $X$ in $T\mathscr{C}_r$ that are orthogonal to $n$: 
\begin{equation}
\label{Xn}
X\cdot n =0;
\end{equation}
but since $X\in H_r$, by definition 
\begin{equation}
\label{Xl}
X\cdot \ell =0.
\end{equation} 
Thus, writing $X=X^r\partial_r+X^t\partial_t+X^i\partial_i$, Eqs. \eqref{Xn} and \eqref{Xl} lead to 
$X^r=0$ and $X^t-b_iX^i=0$, so that 
\begin{equation}
\label{}
X\in H_r \Leftrightarrow X=X^i\left(\partial_i+b_i\partial_t\right)= X^iE_i.
\end{equation} 
Consequently, the field $b_i(r,t,\mathbf{x})$ plays the role of an Ehresmann connection for each null leave $\mathscr{C}_r$, as one could have anticipated. Notice also that the tensor $p^a_{\hphantom{a}b}=\ell^a n_b$ has non-zero components $p^t_{\hphantom{t}t}$ and $p^t_{\hphantom{t}i}$. These define a Carrollian tensor, which is the vertical Ehresmann projector $p$ introduced in \eqref{Ehres}.

Given the above embedding of null hypersurfaces $\mathscr{C}_r$, we can determine their extrinsic geometry. This is generally captured by three quantities: the surface gravity, the deformation tensor and the twist, all built with the projector onto $H_r\subset T\mathscr{C}_r$:
\begin{equation}
h^a_{b}= \delta^a_{b}-n^a \ell_b-\ell^an_b.
\end{equation} 
Lowering an index we find that the non-zero components are $h_{ij}=g_{ij}(r,t,\mathbf{x})$ and the surface gravity vanishes  with our choice of $\ell$. The other extrinsic quantities are
respectively given by
\begin{equation}
\begin{split}
&D^{ab}=\frac{1}{2}h^{ac}h^{bd}\mathcal{L}_{\ell}\ h_{cd},\\
&\omega_a=h^{b}_a n_c\nabla_b\ell^c,
\end{split}
\end{equation}
where $\nabla_a$ stands for the Levi--Civita connection of $g_{ab}$. In addition, the deformation tensor is reduced to the expansion and the shear:
\begin{equation}
\begin{split}
&\Theta=h_{ab}D^{ab}=\frac{1}{2}h^{ab}\mathcal{L}_{\ell}\ h_{ab},\\
&\sigma^{ab}=D^{ab}-\frac{\Theta}{d}h^{ab}.
\end{split}
\end{equation}
For the geometry at hand, the non vanishing components of the extrinsic tensors, at every $r$, coincide with the Carrollian tensors defined on $\mathscr{C}_r$ (see \eqref{varphi}, \eqref{thetashe}):
\begin{equation}
\begin{split}
&\omega_i=-\frac{1}{2} \partial_tb_i-\frac{1}{2\Omega}(\partial_i\Omega+b_i\partial_t\Omega)
=-\frac{1}{2}\varphi_i,\\
&\Theta=\frac{1}{\Omega}\partial_t\ln\sqrt{g}=\theta,\\
&\sigma_{ij}=\frac{1}{2\Omega}\partial_tg_{ij}-\frac{\Theta}{d}g_{ij}=\zeta_{ij}.
\end{split}
\end{equation}
The reduced bulk covariance \eqref{BulkDiff}, which preserves the form \eqref{ambmet}, corresponds precisely to the Carrollian diffeomorphisms \eqref{cardiff}, for which these objects are genuine tensors.
 
In conclusion, before we turn to the investigation of conformal isometries, the message is that the definition of Carrollian spacetimes as fibre bundles with Ehresmann connection and a degenerate metric is adapted to the description of families of embedded null hypersurfaces where, on any leaf, the induced geometry is Carrollian.

\section{Conformal Carrollian Isometries} \label{BMS}

Carrollian spacetimes $\mathscr{C}$ have been introduced in Sec. \ref{carfib} irrespective of any isometry properties. Carrollian diffeomorphisms are not isometries. They are a subgroup of the full diffeomorphism group, compatible with the intrinsic splitting in vertical versus horizontal components of the tangent bundle $T\mathscr{C}$, made possible thanks to the Ehresmann connection. The Carroll group emerges precisely on the tangent space at a point.
Suppose indeed that we trade the $H$ basis vectors $E_i$ for a set of vectors $\hat E_{\hat{\imath}}$, orthonormal with respect to $g$: $g\big(\hat E_{\hat{\imath}},\hat E_{\hat{\jmath}}\big)=\delta_{\hat{\imath}\hat{\jmath}}$. The tangent space is now everywhere spanned by $\{E,\hat E_{\hat{\imath}}, \hat{\imath}=1,\ldots,d\}$, whereas for the cotangent space the basis is $\{\pmb{e},\hat{\pmb e}^{\hat{\imath}}, \hat{\imath}=1,\ldots,d\}$ with $\hat{\pmb e}^{\hat{\imath}}\big(\hat E_{\hat{\jmath}}\big)=\delta_{\hat{\jmath}}^{\hat{\imath}}$. Automorphisms of the tangent space preserving the vertical vector field $E$ and the orthonormal nature of the $H$ basis are generally as follows:
\beq
\begin{pmatrix}
E^{\prime}&\hat E_{\hat{\imath}}^{\prime}
\end{pmatrix}
=
\begin{pmatrix}
E&\hat E_{\hat{\jmath}}
\end{pmatrix}
\begin{pmatrix}
1&B_{\hat{k}} R^{\hat{k}}{}_{{\hat{\imath}}}\\
0&R^{\hat{\jmath}}{}_{{\hat{\imath}}}
\end{pmatrix}
\label{transfo}
\eeq
with $R^{\hat{k}}{}_{\hat{\imath}}(t,\mathbf{x})$ the elements of a $d$-dimensional orthogonal matrix and $B_{\hat{k}}(t,\mathbf{x})$, $d$ numbers. The explicit dependence on the coordinates underlines that this transformation needs not be the same at every point of $\mathscr{C}$.
These transformations are the $d+1$-dimensional Carroll boosts (the full Carroll group also includes spacetime translations). They rotate the horizontal frame and coframe, and produce a rotation plus a shift proportional to ${B}$ on the Ehresmann connection. This latter statement can be made explicit by writing
\begin{equation}
\hat E_{\hat{\imath}}= E^{j}{}_{\hat{\imath}}\partial_j+b_{\hat{\imath}}\partial_t;
\end{equation}
the transformation \eqref{transfo} thus implies
\begin{equation}
E^{\prime j}{}_{\hat{\imath}}=E^{j}{}_{\hat{k}}R^{\hat{k}}{}_{{\hat{\imath}}}\quad\text{and}\quad b^{\prime}_{\hat{\imath}}=\left(b_{\hat{k}}+\Omega^{-1} B_{\hat{k}}\right)R^{\hat{k}}{}_{{\hat{\imath}}}.
\end{equation}
The Carroll boosts play for the tangent bundle of a Carrollian spacetime the same role as the Lorentz group does for the tangent bundle of a pseudo-Riemannian manifold. 

The Carroll group appears also as the isometry group of the flat Carroll manifold introduced in Eqs. \eqref{stflcar}. These isometries are diffeomorphisms generated by vectors $\xi$ such that $\mathcal{L}_\xi g=0$, $\mathcal{L}_\xi E=0$, and shifting the Ehresmann connection by an arbitrary constant. One finds:
\beq\label{stflcar-iso}
\xi^0= \beta_jx^j+ \gamma,\quad \xi^i=\omega^i{}_{j} x^j + \epsilon^i
\eeq
with all entries constant and $\omega_{kj}=\delta_{ki}\omega^i{}_{j}$ antisymmetric. These are precisely the ${(d+2)(d+1)}/{2}$ generators of the Carroll algebra $\mathfrak{carr}(d+1)$.

We would like to enter now the core of our discussion about conformal Carrollian isometries for generic Carrollian spacetimes. We will first define them, and then solve the associated differential equations under the assumption of the absence of shear. This will enable us to exhibit a rather universal algebra, which gives a generalized version of the infinite-dimensional conformal Carroll algebra $\mathfrak{ccarr}(d+1)$.

We define Carrollian conformal Killing vector fields $\xi$ by imposing
\begin{equation}
\mathcal{L}_{\xi}g=\lambda g,
\label{ConfSpace}
\end{equation}
where $\lambda(t,\mathbf{x})$ is an \emph{a priori} arbitrary function. Setting $\xi= f(t,\mathbf{x})E+\xi^i(t,\mathbf{x})E_i$ we obtain:
\begin{eqnarray}
\mathcal{L}_{\xi}g&=& 
\Big(2g_{ij} \partial_t \xi^i\Big) \text{d}t \text{d}x^j
+ \bigg(\left(\Omega^{-1} f+b_k\xi^k\right)\partial_t  g_{ij}
+\xi^k\partial_k g_{ij}+g_{ik}\partial_j \xi^k+g_{jk}\partial_j \xi^k\bigg)\text{d}x^i\text{d}x^j
\label{ConfSpace-eq-nat}
\\
&=&
\Big(2\Omega^{-1}g_{ij} \partial_t \xi^i \Big)\pmb e \pmb e^j+ 
\Big(2 f\left(\zeta_{ij}+\tfrac{1}{d}\theta g_{ij}\right)+D_i \xi_j+D_j \xi_i
\Big)\pmb e^i\pmb e^j,
\label{ConfSpace-eq}
\end{eqnarray}
where $D_i$ stands for the Levi--Civita--Carroll connection introduced in \eqref{dgammaCar}. Observe that the time dependence of the metric enters  these expressions explicitly and one might expect it to alter significantly the structure of the conformal isometry algebra. At the same time one should also stress that {\it in the absence of time dependence}, neither the Ehresmann connection nor the scale factor $\Omega(t,\mathbf{x})$ play a role in the analysis of conformal properties, which would reduce to the analysis in \cite{Duval:2014uoa, Duval:2014uva, Duval:2014lpa}.\footnote{Notice that $\xi=\left(\Omega^{-1}f+b_k\xi^k\right)\partial_t+\xi^i\partial_i$. 
Equation \eqref{ConfSpace-eq-nat} depends on $b_k$ and $\Omega$ only through $ \xi^t\equiv\Omega^{-1}f+b_k\xi^k$.
}
The first term of \eqref{ConfSpace-eq} translates through Eq. \eqref{ConfSpace} into  
\begin{equation}
\partial_t\xi^i(t,\mathbf{x})=0.
\label{ConfTime}
\end{equation}
This imposes that $\xi$ is the generator of a Carrollian diffeomorphism (it ensures the vanishing entry in \eqref{jac} since it imposes $\xi^i(t,\mathbf{x})=Y^i(\mathbf{x})$), and this is assumed systematically here. Hence the core of the definition of conformal Carrollian isometries is in the second term of \eqref{ConfSpace-eq}, leading to
\begin{equation}
2 f\left(\zeta_{ij}+\frac{1}{d}\theta g_{ij}\right)+D_i Y_j+D_j Y_i=\lambda g_{ij}.
\label{ConfSpace2}
\end{equation}
The trace of this equation determines $\lambda$,
\begin{equation}
\lambda(t,\mathbf{x})=\frac{2}{d}\left( f\theta+D_iY^i\right)(t,\mathbf{x}),
\label{Trace}
\end{equation}
and substitution back into \eqref{ConfSpace2} then gives
\begin{equation}
D_i Y_j+D_j Y_i-
\frac{2}{d}D_k Y^k
g_{ij}=-2 f\zeta_{ij}.
\label{ConfSpace2bis}
\end{equation}

At the present stage, the equations to be solved for finding the components of the conformal Killing vectors $ f(t,\mathbf{x})$ and $Y^i(\mathbf{x})$ are Eqs. \eqref{ConfSpace2bis}, which are a set of time-dependent partial differential equations sourced by the Carrollian shear.

In the Carrollian case under consideration, as a consequence of the degenerate nature of the metric, this set -- in other words Eq.~\eqref{ConfSpace} -- is not sufficient for defining conformal Killing fields. In order to proceed, we must refine our definition of the latter. We will further impose vanishing shear for the Carrollian spacetime, and with this the full conformal algebra can be unravelled without any further restriction on the Carrollian data $g_{ij}$, $\Omega$ and $b_i$, generalizing thereby the range of validity of the results obtained in  \cite{Duval:2014uoa, Duval:2014uva, Duval:2014lpa}. 

We note that for $\xi= f(t,\mathbf{x})E+Y^i(\mathbf{x})E_i$, the Lie derivative of the vertical vector field $E$ is itself vertical, satisfying
\begin{equation}
\label{CarrStrDiff}
{\cal L}_{\xi} E=\mu E,
\end{equation}
where
\begin{equation}
\mu(t,\mathbf{x})=-E( f)-\varphi_iY^i.
\label{ConfTime2}
\end{equation}
A precise definition of the conformal Carrollian Killing vectors is reached by setting a relation among the \emph{a priori} independent functions $\lambda(t,\mathbf{x})$ and $\mu(t,\mathbf{x})$. The guideline for this is Weyl covariance, because a desirable feature for conformal Killing fields is their insensitivity to Weyl rescalings of the metric. 

We define Weyl rescalings as $g\mapsto g/{\cal B}(t,\mathbf{x})^2$ and $\pmb b$ invariant (this is required for the spatial vectors $E_i$ in \eqref{Di} to remain well-defined), supplemented with $\Omega(t,\mathbf{x})\mapsto {\cal B}(t,\mathbf{x})^{-z}\Omega(t,\mathbf{x})$ for some real number $z$, the dynamical exponent. Under such rescalings, $\xi$ has Weyl weight zero which implies that $Y^i$ and $f$ have weights zero and $-z$. Therefore $\lambda(t,\mathbf{x})$ and $\mu(t,\mathbf{x})$ 
transform as 
\beq
\lambda\mapsto \lambda-2Y^iE_i(\ln{\cal B}),\quad \mu\mapsto \mu+zY^iE_i(\ln{\cal B}).
\eeq
Thus, the combination $2\mu+z\lambda$ is Weyl covariant (actually invariant). Setting it to zero 
\begin{equation}
2\mu(t,\mathbf{x})+z\lambda(t,\mathbf{x})=0
\label{Constraint}
\end{equation}
is compatible with the basic expected attributes of Killing vectors, as stressed earlier.

Equations \eqref{ConfSpace} and \eqref{Constraint} define our conformal Killing fields. It should be mentioned that \eqref{Constraint} was introduced in \cite{Duval:2014uoa, Duval:2014uva, Duval:2014lpa} with $z=-2/N$ and $N$ a positive integer, following the requirement that ${\cal L}_{\xi} \left(g\otimes E^{\otimes N}\right)=0$. Leaving $z$ arbitrary does not support such a geometrical interpretation, but is nonetheless consistent.  The case $z=1$ (\emph{i.e.}, $N=2$), where time and space equally dilate, pertains when the Carrollian spacetime emerges on an embedded null hypersurface in a pseudo-Riemannian geometry. 

The combination of \eqref{Trace}, \eqref{ConfTime2} and \eqref{Constraint} leads to\footnote{The left-hand side of Eq. \eqref{ConfTime2bis} can actually be recast using Weyl-covariant  derivatives, based on the {\it Weyl connection} $\pmb A=\frac{1}{z}\pmb{\varphi}+\frac{1}{d}\theta \pmb{e}$, which transforms as $\pmb A\mapsto\pmb A-\text{d}\ln{\cal B}$.}
\beq
\label{ConfTime2bis}
D_iY^i-\frac{d}{z}\varphi_iY^i-\frac{d}{z}\left( E( f)-\frac{z}{d}\theta f\right)=0.
\eeq 
Summarizing,  the conformal isometry group as defined in \eqref{ConfSpace} and \eqref{Constraint} for
a Carrollian spacetime described in terms of $\Omega(t,\mathbf{x})$, $b_i(t,\mathbf{x})$ and $g_{ij}(t,\mathbf{x})$ is the set of solutions $f(t,\mathbf{x})$ and $Y^i(\mathbf{x})$ of Eqs.  \eqref{ConfSpace2bis} and (\ref{ConfTime2bis}) for a given choice of $z$. 

At this point we will restrict our analysis to Carroll spacetimes with vanishing shear, $\zeta_{ij}=0$, because in this case the system (\ref{ConfSpace2bis}, \ref{ConfTime2bis})  can be solved. 
As stated previously, $\zeta_{ij}$ vanishes if and only if the time dependence of the metric is conformal:
\beq
\label{confmetric}
g_{ij}(t,\mathbf x)=\text{e}^{2\sigma(t,\mathbf x)}\tilde g_{ij}(\mathbf x).
\eeq
Recall now that (\ref{ConfSpace2bis}, \ref{ConfTime2bis}) are Weyl covariant. Performing a Weyl rescaling with $\mathcal{B}(t,\mathbf{x})=\text{e}^{2\sigma(t,\mathbf x)}$ removes the time-dependence from the metric, while it transforms the other fields as
\beq
\label{fixedOmega}
\tilde\Omega(t,\mathbf{x})=\text{e}^{-z\sigma(t,\mathbf{x})}\Omega(t,\mathbf{x}),\qquad 
\tilde\varphi_i(t,\mathbf{x})=\varphi_i(t,\mathbf{x})-z(\partial_i+b_i(t,\mathbf{x})\partial_t)\sigma(t,\mathbf{x}),\qquad \tilde\theta(t,\mathbf{x})=0.
\eeq
The Killing field is invariant, $ \tilde\xi=\xi= \tilde{f}\tilde{E}+ Y^i E_i $ with $\tilde E =\text{e}^{z\sigma}E $, and this leads to 
\begin{equation}
\tilde f(t,\mathbf{x})=\text{e}^{-z\sigma(t,\mathbf{x})}f(t,\mathbf{x}),\qquad
\tilde Y^i(\mathbf{x})=Y^i(\mathbf{x}), \qquad \tilde{Y}_i(\mathbf{x})=\tilde g_{ij}(\mathbf{x})Y^j(\mathbf{x}).
\end{equation}
Equations \eqref{ConfSpace2bis} and \eqref{ConfTime2bis} finally become equations for $ \tilde f(t,\mathbf{x})$ and $ Y^i(\mathbf{x})$: 
\begin{eqnarray}
\tilde{\nabla}_i{Y}_j+\tilde{\nabla}_j{Y}_i&=&\frac{2}{d}\tilde{\nabla}_kY^k\tilde g_{ij},
\label{ConfEq}\\
\label{matchingthree}
\tilde\Omega^{-1} \partial_t \tilde f&=& \frac{z}{d}\tilde{\nabla}_kY^k-\tilde\varphi_kY^k,
\end{eqnarray}
where $\tilde{\nabla}_i$ is the Levi--Civita connection for $\tilde g_{ij}$.

The first equation is an ordinary conformal Killing equation, and its solutions $\{Y^i(\mathbf{x})\}$ are the generators 
of the conformal group for $\mathscr{S}$ equipped with a metric $\tilde g_{ij}(\mathbf x)$.  Given any such vector in $H$ solving  \eqref{ConfEq}, 
\begin{equation}
\label{XY}
\bar\xi_Y=Y^i(\mathbf x) E_i= Y^i(\mathbf{x})\left(\partial_i +b_i(t,\mathbf{x})\partial_t \right)
\end{equation} 
(the subscript ``$Y$''  stresses that the vector field at hand depends on the set $\{Y^i(\mathbf{x})\}$),
Eq. \eqref{matchingthree} provides a solution for $\tilde f(t,\mathbf{x})$: 
\begin{equation}
\label{fsol}
\tilde{f}(t,\mathbf{x})={T}(\mathbf{x})+\frac{z}{d}\int^t \text{d}{t^*} \; \tilde{\Omega}\left({t^*},\mathbf{x}\right)\left(\tilde{\nabla}_iY^i(\mathbf{x})-\frac{d}{z}\tilde\varphi_i\left({t^*},\mathbf{x}\right)Y^i(\mathbf{x})\right).
\end{equation} 
Here ${T}(\mathbf{x})$ is an arbitrary smooth function of weight $-z$, which specifies any conformal Carrollian Killing field. 

Before we further investigate this family of conformal Carrollian  Killing vectors, we  should pause and make contact with previous results reached in the already quoted literature. The situation that has been studied in \cite{Duval:2014lpa} corresponds in our language to $\sigma=0$ and  $\Omega = 1$.
This means in particular that the metric is time-independent. In Ref. \cite{Duval:2014lpa} no Ehresmann connection was introduced. We could therefore set it to zero, or better  leave $b_i(t,\mathbf{x})$ unspecified, because, as mentioned earlier for a time-independent metric, it is not expected to play any role in the conformal algebra. Indeed, using \eqref{ConfEq} we find the precise family of vectors $\bar\xi_Y$ as in \eqref{XY}, which combined with \eqref{fsol} lead to
\begin{equation}
\label{duval}
\xi_{T,Y}=\left({T}(\mathbf{x})+\frac{z}{d}t \tilde \nabla_i Y^i(\mathbf{x})\right)\partial_t+Y^i(\mathbf x) \partial_i
\end{equation} 
\emph{irrespective of $b_i(t,\mathbf{x})$} (again the subscript ``$T,Y$''  reminds the dependence on $\{T(\mathbf{x}),Y^i(\mathbf{x})\}$). Therefore the corresponding algebra is  infinite-dimensional and emerges as the semi-direct product of the conformal group of $g=\tilde g(\mathbf{x})$  on $\mathscr{S}$, generated by $Y^i(\mathbf x) \partial_i$, with supertranslations. For a flat, or conformally flat metric on $\mathscr{S}$, the spatial conformal algebra in $d$ dimensions is  $\mathfrak{so}(d+1,1)$, and the conformal Carrollian Killing fields \eqref{duval} span\footnote{This algebra is defined in the literature for integer $N$.} $\mathfrak{ccarr}_N(d+1)=\mathfrak{so}(d+1,1)\ltimes \mathfrak{T}_N$, where $z=2/N$. The standard conformal Carrollian algebra $\mathfrak{ccarr}(d+1)$ refers to dynamical exponent $z=1$ (level $N=2$): $\mathfrak{ccarr}(d+1)=\mathfrak{ccarr}_2(d+1)$. This algebra emerges as the null-infinity isometry algebra of asymptotically flat $d+2$-dimensional spacetimes in Bondi gauge, $\mathfrak{bms}(d+2)$.\footnote{As before, strictly speaking this is valid for $d=1$ and $2$ (where furthermore $\tilde g$ is always conformally flat). For higher $d$, it was presumed to hold by some authors \cite{Duval:2014uva}. However, gauge conditions exist for the Bondi-gauge null-infinity behavior of asymptotically flat spacetimes that render $\mathfrak{bms}(d+2)$ finite-dimensional \cite{Campoleoni:2017qot}, and with this choice $\mathfrak{ccarr}_2(d+1)\neq \mathfrak{bms}(d+2)$. This does not exclude that less restrictive gauge fixing might be considered leading to other, possibly infinite-dimensional $\mathfrak{bms}(d+2)$ algebras for $d\geq 3$.}

Our general analysis embraces the above case, by including time dependence in the spatial metric $g$ and a general scale factor $\Omega(t,\mathbf{x})$ on top of the Ehresmann connection $b_i(t,\mathbf{x})$. Despite these generalizations, as a direct consequence of the factorized time dependence in the metric 
(see \eqref{confmetric}) due to the requirement of vanishing shear, the structure of the conformal Carrollian Killing vectors remains unaltered \emph{i.e.}, as in \eqref{duval}: their algebra is the semi-direct product of the conformal group of $\tilde g(\mathbf{x})$ on $\mathscr{S}$ with supertranslations at dynamical exponent $z$.  This statement is shown as follows.

Using \eqref{fsol}, we obtain the general conformal Carrollian Killings as vector fields in $T\mathscr{C}$:
\begin{equation}
\label{genKil}
\xi_{T,Y}=\left({T}(\mathbf{x})+\frac{z}{d}\int^t \text{d}{t^*} \; \tilde{\Omega}\left({t^*},\mathbf{x}\right)\left(\tilde{\nabla}_iY^i(\mathbf{x})-\frac{d}{z}\tilde\varphi_i\left({t^*},\mathbf{x}\right)Y^i(\mathbf{x})\right)\right)\tilde E+Y^i(\mathbf x) E_i.
\end{equation} 
We can unravel the structure of these conformal Carroll Killings and of their algebra by introducing 
an \hlt{invariant local clock}:
\beq
\label{clock}
C(t,\mathbf{x})\equiv \int^t \text{d}{t^*}\; \tilde\Omega\left({t^*},\mathbf{x}\right).
\eeq
This in fact is a specific instance of
$C_\gamma= \int_{\gamma}\tilde \Omega (\text{d}t-b)$
with $\gamma$ a path in $\mathscr{C}$.  In \eqref{clock}, $C(t,\mathbf{x})$ appears as a local function because the path runs along a vertical fibre starting at, say, the zero section, reference to which we have suppressed.\footnote{We refer to $C(t,\mathbf{x})$ as invariant local clock because it defines an integration measure on each one-dimensional fiber, a proper time.}
Using \eqref{varphi} and \eqref{clock} we reach the following identity:
\beq
\int^t \text{d}{t^*}\ \tilde\Omega\left({t^*},\mathbf{x}\right)\tilde\varphi_i\left({t^*},\mathbf{x}\right)=E_i\left(C(t,\mathbf{x})\right),
\eeq
which enables us to express \eqref{genKil} as
\beq
\label{Xclock}
\xi_{{T},Y}=\left({T}(\mathbf{x})
-Y^iE_i\left(C(t,\mathbf{x})\right)+\frac{z}{d}C(t,\mathbf{x})\tilde{\nabla}_iY^i(\mathbf{x})\right){\tilde E}+Y^i(\mathbf{x})E_i
.
\eeq
The invariant clock defines a Carrollian diffeomorphism (see \eqref{cardiff}) with $t'=C(t,\mathbf{x})$ and $\mathbf{x}'=\mathbf{x}$. Under this diffeomorphism $ \tilde\Omega\to 1$, $\tilde E\to \partial_{t'}$, while \eqref{Xclock}  reads now precisely as  \eqref{duval} with $t$ traded for $t'$. This demonstrates the earlier statement about the algebra of conformal Carrollian Killing vectors of a shearless Carroll spacetime. 

Summarizing, shearless Carrollian  spacetimes, \emph{i.e.} spacetimes equipped with a metric of the form $g_{ij}(t,\mathbf x)=\text{e}^{2\sigma(t,\mathbf x)}\tilde g_{ij}(\mathbf x)$, have a conformal isometry algebra that depends only on $\tilde g(\mathbf x)$, $d$ and $z$: it is the semi-direct product of the conformal algebra of  $\mathscr{S}$ equipped with $\tilde g(\mathbf x)$ and supertranslations at level $N=2/z$. This conclusion is valid irrespective of $\Omega(t,\mathbf{x})$ and $b_i(t,\mathbf{x})$. On the one hand, $\Omega(t,\mathbf{x})$ can disappear from the expression \eqref{Xclock} of the Killings upon an appropriate Carrollian diffeomorphism driven by the invariant local clock. Hence its presence does not affect the algebra. On the other hand, although the Ehresmann connection $b_i(t,\mathbf{x})$ cannot be removed with Carrollian diffeomorphisms (unless its field strength $\pmb\varpi$ and acceleration $\pmb\varphi$ vanish), it cancels out between the last two terms 
in  \eqref{Xclock}. This is not insignificant though, and we would like to discuss it in the remaining of the present chapter.

The set of vectors $Y=Y^i(\mathbf{x})\partial_i\in T\mathscr{S}$ with $\{Y^i(\mathbf{x})\}$ solving \eqref{ConfEq} realize the conformal algebra of $\tilde g$:
\begin{equation}
\label{ccar}
\left[Y,Y^\prime\right]=\left[Y^i\partial_i, Y^{\prime j}\partial_j
\right]
= Y^{\prime\prime k}\partial_k=Y^{\prime\prime}
\end{equation}
with 
\begin{equation}
Y^{\prime\prime k}= Y^i\partial_i(Y^{\prime k})- Y^{\prime i}\partial_i(Y^k).
\end{equation}
These vectors act generally on functions $\phi(\mathbf{x})$. One may instead contemplate a realization in terms of Carrollian vectors $\bar\xi_Y\in H$ as in \eqref{XY}
acting on functions $\Phi(t,\mathbf{x})$ of $\mathscr{C}$. In this case, 
\begin{equation}
\label{XYXY}
\left[ \bar\xi_Y,\bar \xi_{Y'}\right] =\bar \xi_{[Y,Y']}-\pmb\varpi( Y, Y') E=\bar \xi_{[Y,Y']}-\tilde {\pmb\varpi}( Y, Y') \tilde E \in  V\oplus H,
\end{equation}
where $\pmb\varpi( Y, Y')=\varpi_{ij}Y^iY^{\prime j}$ and $\tilde{\pmb \varpi}=\text{e}^{-z\sigma}\pmb \varpi$. Because of the Ehresmann connection, this realization is not faithfully the conformal algebra \eqref{ccar} of $\tilde g$, except if the Carrollian torsion is zero (horizontal piece of the Ehresmann curvature), which coincides with the condition for $H$ to be integrable\footnote{Generally, one expects invariants that prevent the horizontal part of the Ehresmann connection from being flat. For example, in $d=2$, one might have non-zero Chern class $c=\frac{1}{2\pi}\int_{\mathscr{S}} \pmb\varpi$.} (or if the action is limited to functions of $\mathbf{x}$ only, which is not what we want). Furthermore the extra $V$-term is not a central extension, unless the Carrollian  acceleration vanishes (in this case $E$ and $E_i$ commute). 

The expression in parentheses present in \eqref{Xclock} suggests to define, for each set 
$\{Y^i(\mathbf{x})\}$ associated with a solution of  \eqref{ConfEq}, a Carrollian operator $M_Y$ acting on any function $\Phi(t,\mathbf{x})$ of $\mathscr{C}$ as
\beq
\label{MY}
M_{ Y}(\Phi)
\equiv Y^i E_i(\Phi)-\frac{z}{d}\Phi\tilde{\nabla}_iY^i.
\eeq
The mapping $Y\to {M}_{Y}$ is a representation of the group of conformal Killing vectors of  $\tilde g$, which however is again not faithful as the commutator exhibits an extra term, similar to the one in \eqref{XYXY}, possibly vanishing in the same circumstances: 
\beq
\label{Mrep}
\left[M_Y,{M}_{Y'}\right](\Phi)\equiv{M}_{ Y}\left({M}_{Y'}(\Phi)\right)-{M}_{ Y'}\left({M}_{ Y}(\Phi)\right)=M_{[ Y, Y']}(\Phi) -\tilde{\pmb{\varpi}}( Y, Y')\tilde E(\Phi).
\eeq 
Using now the map  \eqref{MY} and $ \bar \xi_Y\in H$ given in Eq. \eqref{XY}, the conformal Killing field in $T\mathscr{C}$, Eq.  
\eqref{Xclock}, is recast as
\beq
\label{confCarrsolgen}
\xi_{{T},Y}=\big({T}(\mathbf{x})-{M}_{ Y}(C)(t,\mathbf{x})
\big)\tilde E+\bar\xi_Y.
\eeq
For vanishing $T(\mathbf{x})$, the representation $M_Y$ defines a lift of $ \bar \xi_Y=Y^iE_i \in H \to T\mathscr{C}$ through the map  
\beq
\label{lift}
\bar \xi_Y \mapsto \xi_{0,Y}=\bar \xi_Y-{M}_Y({C})\tilde{E}. 
\eeq
This lift provides a \emph{faithful and Carrollian (\emph{i.e.}, acting on functions of $t$ and $\mathbf{x}$)} realization of the conformal isometry algebra \eqref{ccar} of $\tilde g$   \emph{on $T\mathscr{C}$}, thanks to the cancellation of the extra term appearing in \eqref{XYXY} and \eqref{Mrep}. 
Even though the Ehresmann connection does not appear ultimately in the conformal algebra, when non-vanishing, it adjusts for making compatible the realization of the algebra with Carrollian diffeomorphism invariance. This is yet another of its numerous facets.
For non-vanishing $T(\mathbf{x})$, we obtain the following commutation relations for the complete conformal Carrollian Killing fields \eqref{confCarrsolgen}:\footnote{We use here the identity $\tilde{E}\left({M}_{Y}({C})\right)=\tilde\varphi_iY^i-\frac{z}{d}\tilde{\nabla}_iY^i$.} 
\begin{equation}
\left[ \xi_{{T}, Y},\xi_{{T}', Y'}\right] =\xi_{{M}_{Y}({T}')-{M}_{ Y'}({T}),[ Y, Y']}.
\end{equation}
This is the usual pattern for conformal Carrollian and BMS algebras.

\section{Conclusions}

In this work, we have considered Carrollian geometries from various perspectives: their defining properties, their emergence on embedded null hypersurfaces and their conformal symmetries. We have emphasized the interpretation of Carrollian spacetime as a fiber bundle endowed with an Ehresmann connection. Realized by a one-form field, this connection defines the splitting of the tangent bundle into vertical and horizontal components. The vertical component coincides precisely with the kernel of a degenerate metric, which is the last piece of equipment for a Carroll structure. It is worth stressing that all defining fields (Ehresmann connection, metric and scale factor) have been assumed space \emph{and} time-dependent throughout the paper.

The vertical versus horizontal canonical separation is preserved by the subset of Carrollian diffeomorphisms.  These enable the reduction of spacetime tensors into purely spatial components, the paradigm being Carrollian torsion and acceleration, emerging as reduced components of the Ehresmann curvature.  Other geometric objects can be introduced using the degenerate metric, such as shear and expansion, and even further based on a horizontal connection, which we only alluded to when discussing the Christoffel--Carroll symbols. Investigating the types of connections that can be defined on the full tangent bundle $T\mathscr{C}$ is an interesting subject
that has been discussed in the literature, but remains incomplete and worth pursuing. 

The above ingredients (Ehresmann connection, vertical and horizontal subbundles) arise naturally on null hypersurfaces embedded in Lorentzian spacetimes, and specific tensors such as Carrollian shear, acceleration and expansion are inherited from the ambient geometry. Our analysis was here confined to the instance of genuine null foliations, but can be adapted to the case of boundary null hypersurfaces, such as black-hole horizons or null infinities.

The last element of our investigation concerns symmetries, and more specifically conformal isometries of Carrollian spacetimes. Contrary to pseudo-Riemannian geometries, the definition of (conformal) isometries cannot rely solely on the Killing equation for the metric, because the latter is degenerate. Here we complied with the standard definition of the conformal Carrollian Killing vectors, and additionally restricted our analysis to the case of \emph{shearless} Carrollian structures. Although seemingly innocuous, as time dependence remains general both in the scale factor and in the Ehresmann connection, this limitation is quite severe. Indeed time dependence of the metric is factorized and this ultimately drives us to the standard semi-direct product of the conformal isometry algebra of the metric with supertranslations. This is infinite-dimensional and coincides with
 $\mathfrak{ccarr}_N(d+1)$, for conformally flat spatial metrics.  One thus recovers 
 $\mathfrak{bms}(d+2)$ in $d=1$ and $2$, and possibly in higher dimension with some appropriate definition of the BMS algebra. Our study has the virtue of sustaining the robustness of the format already known to emerge in static Carrollian spacetimes without scale factor or Ehresmann connection. It stresses the role of the shear, but leaves open the probe of the conformal Carrollian isometries, when the latter is non-zero. It also illustrates another subtle role of the Ehresmann connection, which allows to lift without alteration the conformal isometry algebra of the metric from the basis tangent bundle $T\mathscr{S}$ to the Carrollian tangent bundle $T\mathscr{C}$.

Although relatively confined, our investigation touches upon several timely and perhaps deep issues. Conformal symmetries and in particular the BMS algebra are known to appear as the backbone of conserved charges in asymptotically flat spacetimes. Alongside, the role of null hypersurfaces has been appreciated in flat holography, where they are expected to replace the time-like foliations relevant in anti-de Sitter holography. In particular, their symplectic structure should play a significant role in giving an alternative reading of the gravitational degrees of freedom. Clearly, Carrollian spacetimes and their symmetries are the central concepts in all these developments, which  deserve further analysis, possibly in the lines of our current work.

\section*{Acknowledgements}

We would like to thank Glenn Barnich, Laurent Freidel, Kevin Morand and Tassos Petkou for fruitful discussions. This research was supported in part by the ANR-16-CE31-0004 contract \textsl{Black-dS-String},
by Perimeter Institute for Theoretical Physics and by the US Department of Energy under contract DE-SC0015655. Research at Perimeter Institute is supported by the Government of Canada through the Department of Innovation, Science, and Economic Development Canada and by the Province of Ontario through the Ministry of Research, Innovation and Science.  Ecole Polytechnique preprint number: CPHT-RR010.022019.

\end{document}